\documentclass{article}

\usepackage{amsmath, amsthm, amssymb}
\PassOptionsToPackage{sort,numbers,square}{natbib}
\usepackage[final]{nips_2016}

\usepackage{natbibspacing}
\usepackage[utf8]{inputenc} 
\usepackage[T1]{fontenc}    
\usepackage{hyperref}       
\usepackage{url}            
\usepackage{booktabs}       
\usepackage{amsfonts}       
\usepackage{nicefrac}       
\usepackage{microtype}      
\usepackage{algorithm}
\usepackage{algorithmic}
\usepackage{multirow}

\newtheorem{theorem}{Theorem}[section]
\newtheorem{corollary}{Corollary}[section]
\newtheorem{definition}{Definition}[section]

\newcommand{\xiao}{\fontsize{9.5pt}{\baselineskip}\selectfont}

\title{Mutual Information Optimally\\ Local Private Discrete Distribution Estimation}

\author{
  Shaowei Wang, Liusheng Huang, Pengzhan Wang, Yiwen Nie,\\\textbf{Hongli Xu, Wei Yang, Xiang-Yang Li, Chunming Qiao*}\\
  University of Science and Technology of China, China\\
  *State University of New York at Buffalo, USA\\
  \texttt{\{wangsw, pzwang, nyw2016\}@mail.ustc.edu.cn,}\\ \texttt{\{lshuang, xuhongli, qubit, xiangyangli\}@ustc.edu.cn, qiao@computer.org}
}

\begin{document}

\maketitle

\begin{abstract}
Consider statistical learning (e.g. discrete distribution estimation) with local $\epsilon$-differential privacy, which preserves each data provider's privacy locally, we aim to optimize statistical data utility under the privacy constraints. Specifically, we study maximizing mutual information between a provider's data and its private view, and give the exact mutual information bound along with an attainable mechanism: $k$-subset mechanism as results. The mutual information optimal mechanism randomly outputs a size $k$ subset of the original data domain with delicate probability assignment, where $k$ varies with the privacy level $\epsilon$ and the data domain size $d$. After analysing the limitations of existing local private mechanisms from mutual information perspective, we propose an efficient implementation of the $k$-subset mechanism for discrete distribution estimation, and show its optimality guarantees over existing approaches.
\end{abstract}

\section{Introduction}\label{sec:intro}
In the form of aged social surveys or modern mobile crowdsourcing, crowd contributed data has been an essential component facilitating the application of statistical learning. One critical issue in crowd powered statistical learning is the tradeoff between data utility and individual's privacy. To this end, differential privacy \citep{dwork2006differential}\citep{dwork2006calibrating} has emerged as de facto standard of privacy definition with comprehensible constraints and formal privacy guarantee, and achieves plenty of theoretical results or mechanisms characterizing the definition, both in centralized database setting (e.g. in \citep{smith2011privacy}\citep{diakonikolas2015differentially}\citep{hardt2010multiplicative}) and in the local setting \citep{kasiviswanathan2008can} (e.g. in \citep{duchi2013localnips}\citep{hsu2012distributed}\citep{kairouz2014extremal}), where each data provider sanitizes their secret data locally and independently.

Specifically, the statistical data utility bounds under local $\epsilon$-differential privacy have been widely studied, such as mutual information bounds in \citep{mcgregor2010limits}\citep{duchi2013local}\citep{kairouz2016discrete}, hypothesis testing risks in \citep{duchi2013local}\citep{kairouz2016discrete}, distribution estimation error bounds in \citep{duchi2013local}\citep{bassily2015local}. However, most of theoretical bounds or their attainable mechanisms focus on the high privacy region that $\epsilon$ near to $0.0$ (e.g. $\epsilon<1.0$). As contrast, in the practical privacy region that reasonably preserves privacy meanwhile remaining acceptable data utility, the privacy level $\epsilon$ usually range from $0.01$ to $10.0$ \citep[Table 1]{hsu2014differential}.

In this work, we study local private data utilities for full privacy region, mainly focus on mutual information and discrete distribution estimation. Each provider's data is modeled as a no-prior-knowledge variable from a data domain $\mathcal{X}$, under the combinatorial representation of optimal local private mechanism, we transform an arbitrary mechanism to an weight amortized mechanism without loss of mutual information. A convex-like property of mutual information under local privacy is then discovered, showing randomly output with a subset of $\mathcal{X}$ with fixed size $k$ is the optimal mechanism, which is termed $k$-subset mechanism. Exact mutual information bounds are then derived by optimizing the choice of $k$. We also show optimality of $k$-subset mechanism over existing local private mechanisms in the context of discrete distribution estimation.

\textbf{Our contributions.}\,\,\, In local $\epsilon$-differential privacy, for full privacy region, we present the exact bound of mutual information between a provider's data and its private view, where mutual information could be seemed as a general measurement of statistical data utilities. A mechanism that matches the exact bound is proposed as $k$-subset mechanism, which randomly outputs a size $k$ subset of the data domain.

We analyses utilities of existing local private mechanisms, especially state-of-art mechanisms for discrete distribution estimation, show their limitations in practical privacy region. Then, in the context of discrete distribution estimation, we provide an efficient implementation of the $k$-subset mechanism, including a data randomizer that has complexities linear to the domain size and a distribution estimator that has complexities linear to the number of data providers. We also give optimality guarantees of $k$-subset mechanism under the measurement of $l_2$-norm. Finally, we extensively evaluate $k$-subset mechanism, the evaluation results show significant advantages of $k$-subset mechanism over existing mechanisms especially in the intermediate privacy region (e.g. $\log{2} \leq \epsilon \leq \log{(d-1)}$).

\section{Mutual Information}\label{sec:bound}
Consider $n$ data providers, each provider holds a secret value $x_i \in \mathcal{X}$, where $\mathcal{X}=\{X_j\}_{i=1}^d$ is the data domain of size $d$. In the local setting of differential privacy, each data provider locally and independently sanitizes $x_i$ through a local $\epsilon$-differential private mechanism $Q_i$, and obtains a private view $z_i \in \mathcal{Z}$ of $x_i$, where $\mathcal{Z}=\{Z_l\}_{l=1}^{|\mathcal{Z}|}$ is the output alphabets or channel space. The $z_i$ instead of $x_i$ is then published to the untrusted data aggregator, who intends to infer statistics (e.g. discrete distribution) from private views $(z_i)_{i=1}^n$.

Since no prior knowledge is assumed for each provider's data $x_i$, the aggregator models $x_i$ as a random sample from uniform distribution $P_u$ with probability $P_u(X_j)\equiv\frac{1}{d}$. We study non-interactive and non-adaptive statistical inference here, hence the private channel $Q_i\equiv Q$. Denote the conditional probability of output $Z_l$ when the input alphabet is $X_j$ as $Q(Z_l|X_j)$, for $\epsilon > 0.0$ and for any pair of secret values $X_j, X_{j'}\in \mathcal{X}$, we say a mechanism $Q$ satisfies local $\epsilon$-differential privacy if:
\begin{equation}\label{eq:constraits}
\sup_{z\in\mathcal{Z}} \frac{Q(z|X_j)}{Q(z|X_{j'})} \leq e^\epsilon.
\end{equation}
The aggregator observes private view $z_i$ with induced marginal distribution $M_u(z)=\int Q(z|x)dP_u(x)$, we aim to maximize mutual information between $z_i$ and $x_i$. Note that the channel space $\mathcal{Z}$ is unlimited, and is not restricted to the original data domain $\mathcal{X}$, this has made the analyses of mutual information bound under local privacy more challenging than the noisy channel cases.

\subsection{Exact Bound}

The following theorem gives the exact mutual information bound under local privacy.
\begin{theorem}\label{the:maxbound} Let $X$ be a sample drawn according to the uniform distribution $P_u$ that takes $d$ states, and $Z$ be the locally $\epsilon$-differentially private view of $X$, the maximum mutual information between $X$ and $Z$ is as follows:
$$\sup_{Q\in M_\epsilon} I(X;Z) = \text{max}_{k=\lfloor \beta \rfloor}^{\lceil \beta \rceil}\{\frac{k\cdot e^\epsilon \log{\frac{d\cdot e^\epsilon}{k\cdot e^\epsilon+d-k}}+(d-k)  \log{\frac{d}{k\cdot e^\epsilon+d-k}}}{k\cdot e^\epsilon +d-k}\}.$$
Where $\beta=\frac{(\epsilon e^\epsilon - e^\epsilon+1) d}{(e^\epsilon-1)^2}$ and $M_\epsilon$ is the set of mechanisms satisfying local $\epsilon$-differential privacy.
\end{theorem}
For simplicity, we define $I_k = {\frac{k\cdot e^\epsilon \log{\frac{d\cdot e^\epsilon}{k\cdot e^\epsilon+d-k}}+(d-k)  \log{\frac{d}{k\cdot e^\epsilon+d-k}}}{k\cdot e^\epsilon +d-k}}$. Theorem \ref{the:maxbound} is actually derived from $\sup_{Q\in M_\epsilon} I(X;Z) \leq I_\beta$, which would be useful for comprehensibly bounding mutual information, such as in Section \ref{sec:related}.

\subsection{Proof and Mechanism}

\textbf{Combinatorial representation \citep{kairouz2014extremal}.}\,\,\,
Recently, the corner property of optimal local private mechanisms for mutual information is uncovered by \citep{kairouz2014extremal}, this is, for any $X_j,X_{j'}\in \mathcal{X}$ and $z \in \mathcal{Z}$, there is an optimal mechanism $Q$ that $\frac{Q(z|X_j)}{Q(z|X_{j'})}$ equals either $e^\epsilon$ or $e^{-\epsilon}$ or $1$, .

As a result, each condition probability column $[\ Q_{*}(z|X_1),Q_{*}(z|X_2),...,Q_{*}(z|X_d)\ ]$ of an optimal mechanism $Q_{*}$ is expressed as a weighted canonical pattern $s$, which is a column in pattern matrix $S^{(d)}=\{e^\epsilon,1\}^d$ with size $d\times2^d$. For instance, we have
\begin{equation}\label{eq:patterns}
S^{(3)}= \left[
           \begin{array}{cccccccc}
             1    & e^\epsilon & 1 & 1 & e^\epsilon & e^\epsilon & 1 & e^\epsilon \\
             1    & 1 & e^\epsilon & 1 & e^\epsilon & 1 & e^\epsilon & e^\epsilon \\
             1    & 1 & 1 & e^\epsilon & 1 & e^\epsilon & e^\epsilon & e^\epsilon \\
           \end{array}
         \right].
\end{equation}
Since the mutual information is invariant to permutation of condition probability columns, and is invariant to merging or splitting of columns with same pattern, similar to \citep{kairouz2014extremal},we define $Q\in\mathbb{R}^{d\times2^d}$ as the result of $S^{(d)}\in\mathbb{R}^{d\times2^d}$ column-wisely product with weight vector $W^{(d)}\in\mathbb{R}^{2^d}$. We say a private mechanism $Q$ is a valid mechanism if summation of each row in $Q$ equals to $1.0$, as each row of $Q$ should be a probability distribution.

\textbf{Lossless transformation.}\,\,\,
We now present a mutual-information-lossless transformation between an arbitrary mechanism $Q$ and its amortized mechanism $\overline{Q}$. Let $S^{(d,k)} \subseteq S^{(d)}$ denote the set of columns that exactly have $k$ entries of $e^\epsilon$, and $W^{(d,k)}$ denote the corresponding probability weights. Apparently, both the size of the set $S^{(d,k)}$ and $W^{(d,k)}$ are $n \choose k$. We also have $S^{(d)}=\bigsqcup_{k=1}^d S^{(d,k)}$ and $W^{(d)}=\bigsqcup_{k=1}^d W^{(d,k)}$, where symbol $\bigsqcup$ is concatenation operator here, for simplicity of representation, we treat $S^{(d)}$ and $S^{(d,k)}$ as ordered sets. For example, the first column in $S^{(3)}$ of equation (\ref{eq:patterns}) composes $S^{(3,0)}$ and the later $3$ columns compose $S^{(3,1)}$.

By defining the amortized weight $\overline{w}^{(d,k)}$ and amortized weight vector $\overline{W}^{(d,k)}$ of $k$-combination set $S^{(d,k)}$ as follows:
\begin{equation}\label{the:amortization}
\overline{w}^{(d,k)}=\text{sum}(W^{(d,k)})/{n \choose k},\ \  \overline{W}^{(d,k)}= \{\overline{w}^{(d,k)}\}^{{n \choose k}},
\end{equation}
with $\overline{W}^{(d)}=\bigsqcup_{k=1}^d \overline{W}^{(d,k)}$ and $\overline{Q}$ as the result of $\overline{S}^{(d)}$ column-wisely product with weight vector $\overline{W}^{(d)}$, we deduce the following corollary.

\begin{corollary}\label{the:transformation}
Let $X$ be a sample drawn according to the uniform distribution $P_u$ that takes $d$ states, and $Z$ and $\overline{Z}$ be the locally $\epsilon$-differentially private view of $X$ under the mechanism $Q$ and $\overline{Q}$ respectively,
$$I(X;Z)\equiv I(X;\overline{Z}).$$
\end{corollary}

The proof of the above corollary is in Appendix \ref{app:transformation}. This corollary implies the transformation from $W$ to $\overline{W}$ doesn't affect the mutual information.

\textbf{Convex-like property.}\,\,\,
We now deep into the additive structure of mutual information $I(X;\overline{Z})$ and give the following corollary:
\begin{corollary}\label{the:max}
Let $X$ be a sample drawn according to the uniform distribution $P_u$ with $d$ states, mechanism $\overline{Q}$ be the amortized mechanism of an arbitrary mechanism $Q$, and $\overline{Z}$ be the locally $\epsilon$-differentially private view of $X$ under the mechanism $\overline{Q}$ respectively, we have
\begin{equation}\label{eq:combination}
\begin{aligned}
&\ \ \ \ I(X;\overline{Z}) \leq \text{max}_{k=0}^d\{\frac{k\cdot e^\epsilon \log{\frac{d\cdot e^\epsilon}{k\cdot e^\epsilon+d-k}}+(d-k)  \log{\frac{d}{k\cdot e^\epsilon+d-k}}}{k\cdot e^\epsilon +d-k}\}.&\\
\end{aligned}
\end{equation}
\end{corollary}
Denote $k^*$ as the corresponding $k$ that maximizes $I_k={\frac{k\cdot e^\epsilon \log{\frac{d\cdot e^\epsilon}{k\cdot e^\epsilon+d-k}}+(d-k)  \log{\frac{d}{k\cdot e^\epsilon+d-k}}}{k\cdot e^\epsilon +d-k}}$, the bound in Corollary (\ref{the:max}) is achievable when $\overline{Q}$ is a $k^*$-subset mechanism (see Definition \ref{def:kcombination}), which is denoted as $Q^{k^*}$. Indeed, Since $I_{k}$ is concave when real value ${k}$ ranges from $0$ to $d$, the $\beta=\frac{(\epsilon e^\epsilon -e^\epsilon+1) d}{(e^\epsilon-1)^2}$ maximize $I_{\beta}$, hence $k^*=\lfloor\beta\rfloor$ or $k^*=\lceil\beta\rceil$  (see detail in Appendix \ref{app:miss}).

Combining the previous attainability result with Corollary \ref{the:transformation} and \ref{the:max}, the upper bound of mutual information between a data provider's secret data $X$ and its private view $Z$ is given in Theorem \ref{the:maxbound}.

\begin{definition}[$k$-subset mechanism]\label{def:kcombination}
An amortized mechanism $\overline{Q}$ is the $k$-subset mechanism $Q^{k}$, if for any $k'\in[0,d]$, the corresponding weight vector $\overline{W}^{(d)}$ is as follows,
\begin{equation*}
\begin{aligned}
{d \choose k'} \overline{w}^{(d,k')} =
    \begin{cases}
        \frac{d}{k'\cdot e^\epsilon +d-k'}, &if\ \ k'=k \\
        0.0, &if\ \ k'\neq k \\
    \end{cases}.
\end{aligned}
\end{equation*}
Equivalently, the conditional probabilities in $Q^k$ for any $Z \subseteq \mathcal{X}, X \in \mathcal{X}$ is as follows:
\begin{equation}\label{mechanism}
\begin{aligned}
Q^{k}(Z|X) =
    \begin{cases}
        \frac{d e^\epsilon}{k\cdot e^\epsilon +d-k} / {d \choose k}, &if\ \ |Z|=k\ \ and\ \ X \in Z \\
        \frac{d}{k\cdot e^\epsilon +d-k} / {d \choose k}, &if\ \ |Z|=k\ \ and\ \ X \notin Z \\
        0.0, &if\ \ |Z| \neq k \\
    \end{cases}.
\end{aligned}
\end{equation}
\end{definition}

\subsection{Limitation of Existing Mechanisms}
From mutual information perspective, we now analyse existing mechanisms that satisfies local $\epsilon$-differential privacy,
to show their statistical data utility gaps with the $k$-subset mechanism $Q^k$. We focus on two private mechanisms that achieve state-of-art accuracy for discrete distribution estimation (see detail in Section \ref{sec:related}): the binary randomized response mechanism and the multivariate randomized response mechanism, both of which are based on the idea of telling truth with limited probability that was proposed by \citet{warner1965randomized} in $1965$.

\subsubsection{Multivariate Randomized Response Mechanism}
In the multivariate randomized response mechanism $Q^{m}$, the output alphabets $\mathcal{Z}$ is the original input domain $\mathcal{X}$, and the conditional probabilities is defined for any $Z \in \mathcal{X}, X \in \mathcal{X}$ as follows:
\begin{equation}\label{eq:mrr}
\begin{aligned}
Q^{m}(Z|X) =
    \begin{cases}
        \frac{e^\epsilon}{e^\epsilon+d-1}, &if\ \ Z=X \\
        \frac{1}{e^\epsilon+d-1}, &if\ \ Z\neq X \\
    \end{cases}
\end{aligned}
\end{equation}

Actually $Q^{m}$ is  equivalent to $1$-subset mechanism $Q^{1}$, the mutual information ( or channel capacity ) $I(Z;X)$ between private view $Z$ under $Q^m$ and the no-prior-knowledge secret data $X$ is $I_1 = {\frac{e^\epsilon \log{\frac{d\cdot e^\epsilon}{e^\epsilon+d-k}}+(d-1)  \log{\frac{d}{e^\epsilon+d-1}}}{e^\epsilon +d-1}}$. The $Q^m$ is a mutual information optimal mechanism only when $I_1\geq I_2$, while when $\frac{(e^\epsilon-1)^2}{\epsilon e^\epsilon - e^\epsilon+1} \leq \frac{d}{2}$, the mechanism $Q^m$ is unable to fully capture mutual information under local $\epsilon$-differential private constraints.

\subsubsection{Binary Randomized Response Mechanism}
In the binary randomized response mechanism $Q^b$, the secret data $x$ is expressed as a bit map of length $d$ with $i$-th bit indicates $x$ equals $X_i$ or not, the mechanism then flips over each bit with probability $\frac{1}{e^{0.5 \epsilon}+1}$ to obtain the private view $z$.  The output alphabets $\mathcal{Z}$ in $Q^b$ is power set of the original secret data domain $\mathcal{X}$. The conditional probabilities in $Q^b$ is defined for any $Z \subseteq \mathcal{X}, X \in \mathcal{X}$ as follows:
\begin{equation}\label{eq:brr}
\begin{aligned}
Q^{b}(Z|X) =
    \begin{cases}
        \frac{e^{0.5 \epsilon (d-|Z|+1)}}{(e^{0.5 \epsilon}+1)^{d}}, &if\ \ X \in Z \\
        \frac{e^{0.5 \epsilon (d-|Z|-1)}}{(e^{0.5 \epsilon}+1)^{d}}. &if\ \ X \notin Z \\
    \end{cases}
\end{aligned}
\end{equation}

The mutual information $I(Z;X)$ between private view $Z$ under $Q^b$ and the no-prior-knowledge secret data $X$ is :
\begin{equation}\label{eq:brrlimit}
\begin{aligned}
&\ \ \ \ \ I(Z;X) = \sum_{k=0}^{d} {{d \choose k} \frac{e^{0.5 \epsilon (d-k-1)} (k\cdot e^\epsilon +d-k)}{(e^{0.5 \epsilon}+1)^{d}\cdot  d} {I_k}}&\\
&\leq \sum_{k=1}^{d-1} {{d \choose k} \frac{e^{0.5 \epsilon (d-k-1)} (k\cdot e^\epsilon +d-k)}{(e^{0.5 \epsilon}+1)^{d}\cdot d} {I_{k^*}}}\,\,\, \leq (1-\frac{e^{0.5 \epsilon (d-1)}+e^{-0.5 \epsilon}}{(e^{0.5 \epsilon}+1)^{d}}) {I_{k^*}}.&\\
\end{aligned}
\end{equation}
where $I_{k^*}$ is the maximal $I_k$ when $k\in[0,d]$. Since binary randomized response mechanism is an amortized private mechanism, as Corollary \ref{the:max} shows, its mutual information bound or channel capacity is strictly dominated by $k$-subset mechanism.

\section{Discrete Distribution Estimation}\label{sec:mechanism}
With each data provider $i$ holding a secret value $x_i \in \mathcal{X}$, where $\mathcal{X}=\{X_j\}_{i=1}^d$, the truly discrete distribution $\theta$ over $n$ data providers is expressed as $\theta_j = \frac{1}{n}\#\{i: x_i=X_j\}$. Each provider randomizes $x_i$ via a local $\epsilon$-differential private mechanism $Q$ to obtain a private view $z_i$, then publishes $z_i$ to the aggregator, who infers a estimation of the truly distribution $\theta$ from observed private views $(z_i)_{i=1}^n$.

In the previous section, under the metric of mutual information, $k$-subset mechanism parameterized with appropriate $k$ according to privacy level $\epsilon$ and the data domain size $d$ has showed to be the data utility optimal local private mechanism. In this section, under the context of discrete distribution estimation, we implement and analyse $k$-subset mechanism, mainly focus on providing an efficient data randomizer for each data provider and a practical distribution estimator for the aggregator, along with its theoretical performance guarantees.

\subsection{Randomizer}
As defined in Definition \ref{def:kcombination}, $k$-subset mechanism randomly responses with a size $k$ subset $Z$ of the original data domain $\mathcal{X}$. The output alphabets $\mathcal{Z}$ of $k$-subset mechanism is the set of all subsets $Z \subseteq \mathcal{X}$ with size $k$, but directly sampling from $\mathcal{Z}$ with size ${d \choose k}$ is unpractical. By exploiting the symmetric property of conditional probabilities in $k$-subset mechanism, we present an efficient private randomizer equivalent to Definition \ref{def:kcombination} in Algorithm \ref{alg:randomizer}.
\begin{algorithm}[H]
    \caption{$k$-Subset Randomizer}
    \label{alg:randomizer}
    \begin{algorithmic}[1]
        \REQUIRE a value $x \in \mathcal{X}$, a privacy level $\epsilon$, the subset size $k \in [1,d]$.
        \ENSURE a size $k$ subset $z \subseteq \mathcal{X}$ that satisfies local $\epsilon$-differential privacy.
        \STATE Initialize $z$ as an empty set, $z=\{\}$
        \STATE $\mathbf{with\, probability}$ $\frac{k\cdot e^\epsilon}{k\cdot e^\epsilon+d-k}$ $\mathbf{\{}$
        \STATE \qquad Insert $x$ into $z$, $z=z \bigcup \{x\}$
        \STATE \qquad Randomly sampling $k-1$ elements $Y$ from $\mathcal{X}-x$ without replacement
        \STATE \qquad Add elements in $Y$ to $z$, $z=z \bigcup Y$
        \STATE $\mathbf{\}}$
        \STATE $\mathbf{else}$ $\mathbf{\{}$
        \STATE \qquad Randomly sampling $k$ elements $Y$ from $\mathcal{X}-x$ without replacement
        \STATE \qquad Add elements in $Y$ to $z$, $z=z \bigcup Y$
        \STATE $\mathbf{\}}$
        \RETURN $z$
    \end{algorithmic}
\end{algorithm}
\vspace*{-1em}

The core of this randomizer is randomly choosing $k$ or $k-1$ elements from $\mathcal{X}-x$, its computational and memory costs are both $O(d)$ by applying reservoir sampling \citep{vitter1985random}. The randomizer is also equivalent to exponential mechanism \citep{mcsherry2007mechanism} on $\mathcal{Z}$ with utility function $u(z)=-\sum_{a \in z} [\,a\neq x\,]$, and $Q(z|x) \propto exp(u(z))$.

\subsection{Estimator}
A natural distribution estimator for $k$-combination is by recording frequency $f(Z_l)$ of each subset $Z_l \subseteq \mathcal{X}$, then fit model $[f(z)]_{z \subseteq \mathcal{X}} \sim Q^T$ using linear regression or lasso regression \citep{tibshirani1996regression}. Specifically, in linear regression, estimated distribution $\hat{\theta}$ is unbiased given as $(Q Q^T)^{-1} Q [f(z)]_{z \subseteq \mathcal{X}}$, where $(Q Q^T)^{-1} Q$ could be written in a concisely closed form, hence this estimator has $\Omega(d {d \choose k})$ arithmetic operations and $\Omega({d \choose k})$ memory costs.

The previous estimator is only applicable when data domain size $d$ and subset size $k$ is small, and is inefficient for relatively large $d$ and $k$. Here, by remapping $z$ to the original domain $\mathcal{X}$, we present an unbiased estimator without recording ${d \choose k}$-sized frequencies or resorting regression. The estimator maintains frequency $f(X_j)$ for each $X_j \in \mathcal{X}$ instead, upon receiving a private size $k$ subset $z_i$, increases $f(X_j)$ for each $X_j \in z_i$. For simplicity, we denote $g_k$ as $\frac{k e^\epsilon}{k e^\epsilon +d-k}$, which is the hit rate of $X_j$ when secret value is $X_j$, denote $h_k$ as $(\frac{k e^\epsilon}{k e^\epsilon +d-k}\cdot \frac{k-1}{d-1}+\frac{d-k}{k e^\epsilon +d-k}\cdot \frac{k}{d-1})$, which is the hit rate of $X_{j'}$ when secret value is $X_j$ ($j\neq j'$). The expectation of $f(X_j)$ is a simple formula of $\theta_j$ as follows:
\begin{equation}\label{eq:expectf}
E[f(X_j)]=n\cdot\theta_j\cdot g_k + n\cdot(1-\theta_j)\cdot h_k.
\end{equation}
The full process of this estimator is described in Algorithm \ref{alg:estimator}. Its memory costs is linear to the data domain size $d$, number of arithmetic operations needed is $(n\cdot k +d)$.
\begin{algorithm}[H]
    \caption{$k$-Subset Estimator}
    \label{alg:estimator}
    \begin{algorithmic}[1]
        \REQUIRE private views $(z_i)_{i=1}^n$.
        \ENSURE an unbiased estimation $\hat{\theta}$ of the truly distribution $\theta$.
        \STATE Initialize $f(X_j)=0$ for any $X_j \in \mathcal{X}$
        \FOR{$z_i\ \ \text{in}\ \ (z_i)_{i=1}^n$}
            \FOR{$X_a \in z_i$}
                \STATE $f(X_a)= f(X_a)+1$
            \ENDFOR
        \ENDFOR
        \FOR{$j=1$ to $d$}
            \STATE $\hat{\theta_j}=\frac{f(X_j)-n \cdot h_k}{n\cdot(g_k-h_k)}$
        \ENDFOR
        \RETURN $\hat{\theta}=(\hat{\theta}_j)_{j=1}^d$
    \end{algorithmic}
\end{algorithm}

\subsection{Subset Size $k$}
From mutual information perspective, Theorem \ref{the:maxbound} gives optimal choice of subset size $k^*$ that $k^*=\lfloor\frac{(\epsilon e^\epsilon - e^\epsilon+1) d}{(e^\epsilon-1)^2}\rfloor$ or $k^*=\lceil\frac{(\epsilon e^\epsilon - e^\epsilon+1) d}{(e^\epsilon-1)^2}\rceil$. Such $k^*$ is a statistical utility optimal choice in $k$-subset mechanism, but in the context of discrete distribution estimation, the utility is measured by more specific metrics between the truly distribution $\theta$ and the estimated distribution $\hat{\theta}$, such as total variation distance ($l_1$-norm), total variance ($l_2$-norm) or maximum absolute error ($l_\infty$-norm).

Here we focus on the total variance error or squared $l_2$-norm : $E[|\hat{\theta}-\theta|_2^2]=\sum_{j=1}^d {E[|\hat{\theta}_j-\theta_j|^2]}$. Since the estimator in Algorithm \ref{alg:estimator} is unbiased, we have $E[|\hat{\theta}_j-\theta_j|^2=Var[\hat{\theta}_j]$. As random variable $\hat{\theta}_j$ is a transformation of the random variable $f(X_j)$ in line $8$ of Algorithm \ref{alg:estimator}, and $f(X_j)$ is sum of binomial random variable $B(n\cdot \theta_j, g_k)$ and $B(n-n\cdot \theta_j, h_k)$, the variance of $\hat{\theta}_j$ is $\frac{\theta_j g_k(1-g_k)+(1-\theta_j)h_k(1-h_k)}{n(g_k-h_k)^2}$, hence the total variance error is as follows:
\begin{equation*}\label{eq:l2error}
E[|\hat{\theta}-\theta|_2^2]=\sum_{j=1}^d{\frac{\theta_j g_k(1-g_k)+(1-\theta_j)h_k(1-h_k)}{n(g_k-h_k)^2}}=\frac{ g_k(1-g_k)+(d-1)h_k(1-h_k)}{n(g_k-h_k)^2}.
\end{equation*}
The total variance error in $k$-subset mechanism is independent of the truly distribution $\theta$ but is related to $k$. The optimal choice of $k$ is given in the following theorem (see detail in Appendix \ref{app:l2k}).

\begin{theorem}\label{the:l2k}
For discrete distribution estimation with the measurement of total variation distance $E[|\hat{\theta}-\theta|_2^2]$, the optimal subset size in $k$-subset mechanism is $\lfloor\frac{d}{1+e^\epsilon}\rfloor$ or $\lceil\frac{d}{1+e^\epsilon}\rceil$. This is, we have:
\begin{equation*}
\begin{aligned}
&\min_{k \in [0,n]} & E[|\hat{\theta}-\theta|_2^2] =\min_{k \in [\lfloor\frac{d}{1+e^\epsilon}\rfloor,\lceil\frac{d}{1+e^\epsilon}\rceil]} & \frac{ g_k(1-g_k)+(d-1)h_k(1-h_k)}{n(g_k-h_k)^2}.\\
\end{aligned}
\end{equation*}
\end{theorem}

More generally, for any private mechanism that the output alphabets is the power set of the data domain, its total variance error bound is dominated by $k$-subset mechanism, since the mechanism could be seemed as a hybrid of $k$-subset mechanism with different $k$. Formally, we give the optimality guarantee of $k$-subset mechanism in follows (see Appendix \ref{app:l2koptimal} for proof):

\begin{theorem}\label{the:l2koptimal}
For any locally $\epsilon$-differentially private mechanism $Q$ that the output alphabets $\mathcal{Z}$ is the power set of the data domain $\mathcal{X}$, if $Q(Z|X) = C_{|Z|}\cdot Q(Z'|X)=C_{|Z|}\cdot Q(Z|X')$ holds for any $X,X' \in \mathcal{X},\, Z,Z' \in \mathcal{Z}$ when $X \in Z$, $X \notin Z'$, $X' \notin Z$ and $|Z|=|Z'|$, where $C_{|Z|}\geq 1$ is a constant depends only on the size of $Z$, using the remapping based estimator as in Algorithm \ref{alg:estimator} to infer a estimation $\hat{\theta}$ of the truly distribution $\theta$, we have:
\begin{equation*}
\begin{aligned}
&\sup_{\theta}\, E[|\hat{\theta}-\theta|_2^2] \geq \min_{k \in [\lfloor\frac{d}{1+e^\epsilon}\rfloor,\lceil\frac{d}{1+e^\epsilon}\rceil]} & \frac{ g_k(1-g_k)+(d-1)h_k(1-h_k)}{n(g_k-h_k)^2}.\\
\end{aligned}
\end{equation*}
\end{theorem}

Theorem \ref{the:l2koptimal} for squared $l_2$-norm is intrinsically similar to the Theorem \ref{the:maxbound} for mutual information, implies that resorting to power set of the original data domain as output alphabets is unnecessary, and responding with a subset of the original data domain with fixed size $k$ gives optimal results under various statistical utility measurements, such as mutual information and $l_2$-norm.

\section{Related Work}\label{sec:related}
\vspace*{-0.7em}
\textbf{Discrete distribution estimation.}\,\,\,
Numerous mechanisms achieve local $\epsilon$-differential privacy, the oldest among them dates back to $1965$ by \citet{warner1965randomized}, and is termed "randomized response". Randomized response and its mutant are still basic building blocks for many local private mechanisms, such as multivariate randomized response \citep{agrawal2005privacy}\citep{mcsherry2007mechanism}\citep{kairouz2014extremal}\citep{kairouz2016discrete} for multiple options, binary randomized response \citep{duchi2013localnips}\citep{duchi2013local}\citep{erlingsson2014rappor} on bit maps, randomized $1$-bit response on random binary matrix \citep{bassily2015local}\citep{chen2016private}. Among them, in the high privacy region (e.g. $\epsilon<1.0$), the binary randomized response \citep{duchi2013local} and the randomized $1$-bit response \citep{bassily2015local} achieves optimal error bounds for distribution estimation and succinct histogram estimation respectively. Some other mechanisms (e.g. O-RR in \citep{kairouz2016discrete}, multi cohorts RAPPOR in \citep{erlingsson2014rappor}, bi-parties mechanism in \citep{wang2016private}) could be seemed as mixture of binary randomized response and multivariate randomized response. Specifically, the O-RR mechanism in \citep{kairouz2016discrete} could simulate $k$-subset mechanism under certain parameters, but it evolves with $\Theta(n)$ computationally expensive hashing, its parameter selection and estimation performance are also quite experimental.

This work summarizes and characterizes these mechanisms in Theorem \ref{the:l2koptimal}. Specifically, multivariate randomized response is equivalent to $1$-subset mechanism, randomized $1$-bit response is expectedly equivalent to $\frac{d}{2}$-subset mechanism, binary randomized response is a hybrid $k$-subset mechanism with $k$ acroses $0$ to $d$. Furthermore, for an arbitrary privacy level $\epsilon$, Theorem \ref{the:l2koptimal} implies that the distribution estimation performance of any hybrid $k$-subset mechanism or $k$-subset mechanism with fixed $k$ is dominated by $k$-subset mechanism when $k$ is around $\frac{d}{1+e^\epsilon}$ , which varies with privacy budget $\epsilon$ and data domain size $d$.

We notice that there are some other local private mechanisms by adding noises (e.g. Laplace noises \citep{dwork2006calibrating}, two-sided geometric noises \citep{ghosh2009universally}\citep{geng2015staircase}) on bit maps, but their mutual information bounds or distribution estimation performances are dominated by their binary version: binary randomized response.

\textbf{Mutual information bound.}\,\,\,
\citet{mcgregor2010limits} studies mutual information bounds and communication complexity under local privacy, and gives bound $I(Z;X_b)\leq \frac{3\epsilon^2}{2}$ for a uniform Bernoulli variable $X_b$. Theorem \ref{the:maxbound} from this work further provides exact bound $I_1=(\frac{\epsilon e^\epsilon}{e^\epsilon+1}-\log{\frac{e^\epsilon+1}{2}})$, which implies $I(Z;X_b)\leq \frac{\epsilon^2}{8}$. Our results can also be easily generalized to the uniform multinoulli variable $X_d$ that taking $d$ states. Specifically, $I(Z;X_d)\leq I_\beta \leq \log{(\frac{e^\epsilon-1}{\epsilon})}+\frac{\epsilon}{e^\epsilon-1}-1$, which concludes $I(Z;X_d)\leq \frac{\epsilon^2}{8}$ for any $d$.

\vspace*{-0.3em}
\section{Simulation Results}\label{sec:exp}
\vspace*{-0.7em}
Extensive experiments are conducted to evaluate the performance of $k$-subset mechanism, with comparison to state-of-art mechanisms: Binary Randomized Response (\textbf{BRR}), Multivariate Randomized Response (\textbf{MRR}), under the error measurements of both $l_2$-norm $\|\hat{\theta}-\theta\|_2^2$ and $l_1$-norm $\|\hat{\theta}-\theta\|_1$. The $k$-subset mechanism with mutual information optimal $k^*$ (in Theorem \ref{the:maxbound}) is denoted as $\mathbf{k^*}$\textbf{-SS}, and the mechanism with $l_2$-norm optimal $k^\#$ (in Theorem \ref{the:l2koptimal}) is denoted as $\mathbf{k^\#}$\textbf{-SS}.

In our experiments, $n=10000$ data providers (participants) are simulated, the privacy level $\epsilon$ range from $0.01$ to $5.0$ and the data domain size $d$ range from $2$ to $256$. During each simulation, the truly distribution $\theta$ is generated randomly, the estimated distribution $\hat{\theta}$ is postprocessed by projecting onto probability simplex \citep{wang2013projection}\citep{duchi2008efficient}. Numerical results are mean error of $100$ repeated simulations.

As the simulation results in Table \ref{tab:result} demonstrated, $k$-subset mechanism outperforms BRR and MRR for arbitrary domain size $d$ and privacy level $\epsilon$, which echoes theoretical results in Theorem \ref{the:l2koptimal}. Especially in intermediate privacy region ( e.g. $\log{2} \leq \epsilon \leq \log{(d-1)}$ or $1< k^\# \leq \frac{d}{3}$ in the table), there are averagely about $20\%$ $l_2$-error reduction and $10\%$ $l_1$-error reduction. The $l_2$-norm optimal mechanism $k^\#$-SS achieves slightly better performance than the mutual information optimal mechanism $k^*$-SS, but the differences are pretty minor.

\begin{table}[H]
\setlength\tabcolsep{0.81ex}
\renewcommand{\arraystretch}{1.10}
\caption{Simulation results of mean $\|\hat{\theta}-\theta\|_2^2$ and $\|\hat{\theta}-\theta\|_1$ error for discrete distribution estimation.}
\small
\centering
\renewcommand{\multirowsetup}{\centering}
\label{tab:result}
\begin{tabular}{|c|c||c|c|c|c||c|c|c|c||c|c|}
\cline{3-10}

\multicolumn{2}{c}{}  & \multicolumn{4}{|c||}{\large{$\mathbf{\|\hat{\theta}-\theta\|_2^2}$}} & \multicolumn{4}{c}{\large{$\mathbf{\|\hat{\theta}-\theta\|_1}$}} & \multicolumn{2}{|c}{} \\
\cline{1-12}

$d=$& $\epsilon=$  & \textbf{BRR} & \textbf{MRR} & $\mathbf{k^*}$\textbf{-SS} & $\mathbf{k^\#}$\textbf{-SS}& \textbf{BRR} & \textbf{MRR} & $\mathbf{k^*}$\textbf{-SS} & $\mathbf{k^\#}$\textbf{-SS} & $\mathbf{k^*}$ & $\mathbf{k^\#}$ \\
\hline

\multicolumn{1}{|c|}{\multirow{2}{*}{2}} & $0.1$ &  0.03361	& 0.009983	& 0.01001	& \textbf{0.009026} 	&  0.2058	& 0.1123	& 0.1132	& \textbf{0.1048} 	& 1 	& 1 \\
\cline{2-12}
& $1.0$ &  0.00043	& 9.75e-05	& 9.82e-05	& \textbf{8.21e-05} 	&  0.02341	& 0.01112	& 0.01084	& \textbf{0.01024} 	& 1 	& 1 \\
\cline{1-12}

\multicolumn{1}{|c|}{\multirow{4}{*}{4}} & $0.01$ &  0.6492	& 0.604	& \textbf{0.5797}	& 0.5822 	&  1.313	& 1.271	& \textbf{1.233}	& 1.242 	& 2 	& 2 \\
\cline{2-12}
& $0.1$ &  0.09319	& 0.06928	& 0.05542	& \textbf{0.05383} 	&  0.4863	& 0.4194	& 0.3728	& \textbf{0.3683} 	& 2 	& 2 \\
\cline{2-12}
& $0.5$ &  0.00427	& 0.00286	& \textbf{0.00262}	& 0.00276 	&  0.1044	& 0.08596	& \textbf{0.08118}	& 0.08458 	& 2 	& 2 \\
\cline{2-12}
& $1.0$ &  0.0011	& \textbf{0.00052}	& 0.00056	& 0.00056 	&  0.05362	& \textbf{0.03664}	& 0.03828	& 0.03787 	& 1 	& 1 \\
\cline{1-12}

\multicolumn{1}{|c|}{\multirow{4}{*}{6}} & $0.01$ &  0.6523	& 0.6956	& 0.6366	& \textbf{0.6356} 	&  1.47	& 1.51	& \textbf{1.45}	& \textbf{1.45} 	& 3 	& 3 \\
\cline{2-12}
& $0.1$ &  0.1195	& 0.1333	& \textbf{0.09006}	& 0.09741 	&  0.6773	& 0.7154	& \textbf{0.5851}	& 0.6112 	& 3 	& 3 \\
\cline{2-12}
& $0.5$ &  0.00740	& 0.00689	& 0.00520	& \textbf{0.00512} 	&  0.1664	& 0.162	& 0.14	& \textbf{0.1398} 	& 3 	& 2 \\
\cline{2-12}
& $1.0$ &  0.00185	& 0.00138	& 0.001216	& \textbf{0.00119} 	&  0.08388	& 0.0723	& 0.06813	& \textbf{0.06764} 	& 2 	& 2 \\
\cline{1-12}

\multicolumn{1}{|c|}{\multirow{5}{*}{8}} & $0.01$ &  0.7177	& 0.7516	& \textbf{0.6477}	& 0.6769 	&  1.613	& 1.633	& \textbf{1.558}	& 1.586 	& 4 	& 4 \\
\cline{2-12}
& $0.1$ &  0.1367	& 0.184	& 0.1186	& \textbf{0.1176} 	&  0.831	& 0.9464	& 0.7708	& \textbf{0.7654} 	& 4 	& 4 \\
\cline{2-12}
& $0.5$ &  0.01015	& 0.01189	& 0.00797	& \textbf{0.00786} 	&  0.2272	& 0.2457	& 0.2003	& \textbf{0.1994} 	& 3 	& 3 \\
\cline{2-12}
& $1.0$ &  0.0025	& 0.00241	& 0.00195	& \textbf{0.00190} 	&  0.1124	& 0.1103	& 0.09971	& \textbf{0.09808} 	& 3 	& 2 \\
\cline{2-12}
&$2.0$ 	&  0.00062	& 0.00032	& 0.0004	& \textbf{0.00031} 	&  0.05651	& 0.04001	& 0.04565	& \textbf{0.03976} 	& 2 	& 1 \\
\cline{1-12}

\multicolumn{1}{|c|}{\multirow{6}{*}{16}} & $0.01$ &  0.7008	& 0.8152	& \textbf{0.6988}	& 0.7109 	&  1.77	& 1.819	& \textbf{1.767}	& 1.777 	& 8 	& 8 \\
\cline{2-12}
& $0.1$ &  0.1593	& 0.2754	& 0.1535	& \textbf{0.1511} 	&  1.188	& 1.431	& 1.178	& \textbf{1.166} 	& 8 	& 8 \\
\cline{2-12}
& $0.5$ &  0.01976	& 0.03955	& 0.017	& \textbf{0.01677} 	&  0.4442	& 0.6302	& 0.4127	& \textbf{0.4097} 	& 7 	& 6 \\
\cline{2-12}
& $1.0$ &  0.00531	& 0.00837	& 0.00447	& \textbf{0.00434} 	&  0.2304	& 0.2896	& 0.2117	& \textbf{0.2086} 	& 5 	& 4 \\
\cline{2-12}
&$2.0$ 	&  0.00132	& 0.00094	& 0.00091	& \textbf{0.00085} 	&  0.1157	& 0.09716	& 0.09577	& \textbf{0.09265} 	& 3 	& 2 \\
\cline{2-12}
&$3.0$ 	&  0.00055	& \textbf{0.0002}	& 0.000295	& 0.00021 	&  0.07454	& \textbf{0.04555}	& 0.05492	& 0.04567 	& 2 	& 1 \\
\cline{1-12}

\multicolumn{1}{|c|}{\multirow{6}{*}{32}} & $0.01$ &  0.7025	& 0.8804	& \textbf{0.6925}	& 0.6969 	&  1.878	& 1.919	& 1.879	& \textbf{1.875} 	& 16 	& 16\\
\cline{2-12}
& $0.1$ &  0.161	& 0.3684	& \textbf{0.1555}	& 0.1565 	&  1.484	& 1.739	& \textbf{1.469}	& 1.487 	& 15 	& 15 \\
\cline{2-12}

& $1.0$ &  0.00971	& 0.02396	& 0.00899	& \textbf{0.00876} 	&  0.4393	& 0.6923	& 0.4256	& \textbf{0.418} 	& 11 	& 9 \\
\cline{2-12}
& $1.5$ &  0.00473	& 0.00822	& 0.004	& \textbf{0.00384} 	&  0.3074	& 0.4042	& 0.2838	& \textbf{0.2779} 	& 9 	& 6 \\
\cline{2-12}
&$2.0$ 	&  0.00261	& 0.00308	& 0.00211	& \textbf{0.00188} 	&  0.2285	& 0.2476	& 0.2059	& \textbf{0.1954} 	& 7 	& 4 \\
\cline{2-12}
&$3.0$ 	&  0.0011	& 0.00056	& 0.00074	& \textbf{0.00055} 	&  0.1495	& 0.1053	& 0.1221	& \textbf{0.1051} 	& 4 	& 2 \\
\cline{1-12}

\multicolumn{1}{|c|}{\multirow{7}{*}{64}} & $0.1$ &  0.1543	& 0.4348	& 0.1536	& \textbf{0.1519} 	&  1.691	& 1.88	& \textbf{1.69}	& \textbf{1.69} 	& 31 	& 30 \\
\cline{2-12}
& $0.5$ &  0.03388	& 0.105	& \textbf{0.03359}	& 0.03383 	&  1.114	& 1.578	& \textbf{1.107}	& 1.11 	& 27 	& 24 \\
\cline{2-12}
& $1.0$ &  0.01476	& 0.04257	& 0.01414	& \textbf{0.01383} 	&  0.7649	& 1.213	& 0.7439	& \textbf{0.7397} 	& 22 	& 17 \\
\cline{2-12}
& $1.5$ &  0.00789	& 0.0193	& 0.0072	& \textbf{0.0068} 	&  0.5603	& 0.8649	& 0.5343	& \textbf{0.5209} 	& 17 	& 12 \\
\cline{2-12}
&$2.0$ 	&  0.00476	& 0.00882	& 0.00403	& \textbf{0.00368} 	&  0.4358	& 0.5903	& 0.3998	& \textbf{0.3823} 	& 13 	& 8 \\
\cline{2-12}
&$3.0$ 	&  0.00206	& 0.00162	& 0.00141	& \textbf{0.00113} 	&  0.2872	& 0.2538	& 0.2382	& \textbf{0.212} 	& 7 	& 3 \\
\cline{2-12}
&$5.0$ 	&  0.00058	& 9.97e-05	& 0.000189	& \textbf{9.95e-05} 	&  0.1523	& 0.06337	& 0.0873	& \textbf{0.06334} 	& 2 	& 1 \\
\cline{1-12}

\multicolumn{1}{|c|}{\multirow{4}{*}{$2^7$}} & $0.1$ &  0.148	& 0.523	& \textbf{0.1431}	& 0.1456 	&  1.824	& 1.95	& \textbf{1.823}	& 1.824 	& 62 	& 61 \\
\cline{2-12}
& $1.0$ &  0.01723	& 0.05896	& 0.01675	& \textbf{0.01658} 	&  1.122	& 1.61	& 1.108	& \textbf{1.103} 	& 43 	& 34 \\
\cline{2-12}
&$3.0$ 	&  0.00358	& 0.00436	& 0.00263	& \textbf{0.00222} 	&  0.5339	& 0.5858	& 0.4573	& \textbf{0.4203} 	& 14 	& 6 \\
\cline{2-12}
&$5.0$ 	&  0.0011	& 0.00025	& 0.00047	& \textbf{0.00024} 	&  0.2964	& \textbf{0.1379}	& 0.1934	& 0.1385 	& 4 	& 1 \\
\cline{1-12}

\multicolumn{1}{|c|}{\multirow{3}{*}{$2^8$}} & $1.0$ &   0.01753	& 0.07743	& 0.01726	& \textbf{0.01703} 	&  1.432	& 1.83	& 1.424	& \textbf{1.417} 	& 87 	& 69 \\
\cline{2-12}
&$3.0$ 	&  0.0049	& 0.00826	& 0.004	& \textbf{0.00345} 	&  0.8757	& 1.093	& 0.7928	& \textbf{0.7389} 	& 29 	& 12 \\
\cline{2-12}
&$5.0$ 	&  0.00187	& 0.000599	& 0.000863	& \textbf{0.00055} 	&  0.5447	& 0.3074	& 0.3707	& \textbf{0.2944} 	& 7 	& 2 \\
\hline
\end{tabular}
\end{table}

\vspace*{-0.3em}
\section{Conclusion}\label{sec:conclusion}
\vspace*{-0.8em}
We study statistical data utility optimal mechanisms and discrete distribution estimation under local $\epsilon$-differential privacy. We firstly provide exact mutual information bound under local privacy for full region of privacy level $\epsilon$. A private mechanism that matches the bound is discovered, the $k$-subset mechanism, which outputs a subset of the original data domain with fixed set size $k$, the optimal $k$ that maximizes mutual information is a formula of the privacy level $\epsilon$ and the data domain size $d$. We then analyse suboptimality of existing local private mechanisms from mutual information perspective. Efficient implementation of $k$-subset mechanism for distribution estimation is proposed, and finally we show optimality of $k$-subset mechanism under the measurement of squared $l_2$ norm.

Our theoretical bounds and mechanisms cover full privacy region, and hence fills the gap between theoretical results and mechanisms (e.g. in \citep{mcgregor2010limits}\citep{duchi2013localnips}\citep{bassily2015local}) that applicable only in the high privacy region (e.g. $0 \leq \epsilon \leq 1$) and practical usable mechanisms.  Especially in the practical using privacy levels (e.g. $\log{2} \leq \epsilon \leq \log{(d-1)}$), numerical results show the $k$-subset mechanism significantly outperforms existing outperforms existing mechanisms for discrete distribution estimation.

\textbf{Limitation and future work.}\,\,\,
An important research direction is the universal mutual information bound ( or channel capacity ) under local privacy, which includes analysing whether the mutual information bounds in Theorem \ref{the:maxbound} hold for $X$ with prior knowledge, and seeking for the optimal domain size $d$ that maximizes mutual information.
\vspace*{-0em}



\xiao
\vspace*{-0em}
\bibliographystyle{plainnat}
\bibliography{refs}
\normalsize

\newpage
\section*{\centering{\emph{Supplementary Materials}}}
\appendix

\section{Proof of Corollary \ref{the:transformation}}\label{app:transformation}
Before proving $I(X;Z)\equiv I(X;\overline{Z})$, we should firstly show $\overline{Q}$ is a valid randomization mechanism. Since $Q$ is a valid mechanism, the total mass of induced marginal probabilities of random variable $Z$ is $1.0$, this is, $\sum_{k=0}^d \frac{1}{d} \text{sum}(W^{(d,k)})(k\cdot e^\epsilon+d-k)=1$. As a result, each row of the amortized mechanism $\overline{Q}$ is a valid probability distribution, since each probability mass nonnegative and the total mass equals $1.0$ as follows:
\begin{equation}\label{eq:mass}
\begin{aligned}
&\ \ \ \ \sum_{k=0}^d ({{n-1 \choose k-1}e^\epsilon+{n-1 \choose k}})\cdot \text{sum}(W^{(d,k)})/{n \choose k}& \\
&=\sum_{k=0}^d (\frac{k}{d}e^\epsilon+\frac{d-k}{d})\cdot \text{sum}(W^{(d,k)}) = 1.&\\
\end{aligned}
\end{equation}

We now proceed to prove $I(X;Z)\equiv I(X;\overline{Z})$. Note that partial mutual information is invariant to permutation of rows, hence the columns that has same number of $e^\epsilon$ have same structure of partial mutual information. We have:
\begin{equation*}
\begin{aligned}
&\ \ \ \ I(X;Z)=\sum_{k=0}^d \text{sum}(W^{(d,k)}){(\frac{k}{d}\cdot e^\epsilon \log{\frac{d\cdot e^\epsilon}{k\cdot e^\epsilon+d-k}}+\frac{d-k}{d}  \log{\frac{d}{k\cdot e^\epsilon+d-k}})} &\\
&= \sum_{k=0}^d (\frac{k}{d}\cdot e^\epsilon +\frac{d-k}{d}) \text{sum}(W^{(d,k)})I_k.&\\
\end{aligned}
\end{equation*}
Similarly, we have:
\begin{equation}\label{eq:amortizedmi}
\begin{aligned}
&I(X;\overline{Z})=\sum_{k=0}^d {({n-1 \choose k-1}\cdot e^\epsilon \log{\frac{d\cdot e^\epsilon}{k\cdot e^\epsilon+d-k}}+{n-1 \choose k}  \log{\frac{d}{k\cdot e^\epsilon+d-k}})}\cdot \text{sum}(W^{(d,k)})/{n \choose k}&\\
&= \sum_{k=0}^d (\frac{k}{d}\cdot e^\epsilon +\frac{d-k}{d}) \text{sum}(W^{(d,k)})I_k.&\\
\end{aligned}
\end{equation}
Combining previous two equations concludes $I(X;Z)\equiv I(X;\overline{Z})$.

\section{Proof of Corollary \ref{the:max}}\label{app:max}
Refer to equation \ref{eq:mass} and \ref{eq:amortizedmi} in Appendix \ref{app:transformation}, we have:
\begin{equation}\label{eq:maxmi}
\begin{aligned}
&\ \ \ \ I(X;\overline{Z})= \sum_{k=0}^d (\frac{k}{d}\cdot e^\epsilon +\frac{d-k}{d}) \text{sum}(W^{(d,k)})I_k &\\
&\leq \sum_{k=0}^d (\frac{k}{d}\cdot e^\epsilon +\frac{d-k}{d}) \text{sum}(W^{(d,k)}) \text{max}_{k'=0}^d \{I_{k'}\} &\\
&\leq \text{max}_{k'=0}^d \{I_{k'}\} &\\
\end{aligned}
\end{equation}
As a result, Corollary $\ref{the:max}$ holds.

\section{Proof of the mutual information optimal subset size $k$}\label{app:miss}
Refer to equation \ref{eq:maxmi}, then denote $k^*$ as the corresponding $k$ that maximizes $I_k$, the equality condition in equation \ref{eq:maxmi} holds when $(\frac{k^*}{d}\cdot e^\epsilon +\frac{d-k^*}{d}) \text{sum}(W^{(d,k^*)})=1.0$ and $\text{sum}(W^{(d,k)})=0.0$ when $k\neq k^*$. We here derive concrete value of $k^*$.

Consider a real value $k \in (0,d)$ as the variable, and the mutual information $I(k)=I_k={\frac{k\cdot e^\epsilon \log{\frac{d\cdot e^\epsilon}{k\cdot e^\epsilon+d-k}}+(d-k)  \log{\frac{d}{k\cdot e^\epsilon+d-k}}}{k\cdot e^\epsilon +d-k}}$ as the objective, we seek for a $k$ that maximizes $I(k)$. The derivative of $I(k)$ is as follows:
\begin{equation*}
\begin{aligned}
I'(k)= \frac{\epsilon e^\epsilon d-(e^\epsilon-1)(k\cdot e^\epsilon+d-k)}{(k\cdot e^\epsilon+d-k)^2}.
\end{aligned}
\end{equation*}
Let $I'(\beta)=0$, we have ${\epsilon e^\epsilon d-(e^\epsilon-1)(\beta\cdot e^\epsilon+d-\beta)}=0$, hence $\beta=\frac{(\epsilon e^\epsilon - e^\epsilon+1) d}{(e^\epsilon-1)^2}$ is the only extrema point in $(0,d)$.
Besides, $I'(k)\geq 0$ holds for $k \in (0,\beta)$, $I'(k)\leq 0$ holds for $k \in (\beta,d)$. In concise, $k=\beta=\frac{(\epsilon e^\epsilon - e^\epsilon+1) d}{(e^\epsilon-1)^2}$ maximizes $I(k)$. Further restricts $k$ to an integer value between $[0,d]$, the mutual information optimal subset size $k=\lfloor\beta\rfloor$ or $k=\lceil\beta\rceil$ holds.

\section{Proof of Theorem \ref{the:l2k}}\label{app:l2k}
Consider a real value $k \in (0,d)$ as the variable, and the total variation distance $f(k)=\frac{ g_k(1-g_k)+(d-1)h_k(1-h_k)}{n(g_k-h_k)^2}$ as the objective, we seek for a $k$ that minimizes $f(k)$. The derivative of $f(k)$ is as follows:
\begin{equation*}
f'(k) = \frac{(d-1)^2((d-k)^2-e^{2\epsilon}k^2)}{(e^\epsilon-1)^2(d-k)k}.
\end{equation*}
Let $f'(k)=0$, we have $d-k=\pm e^\epsilon k$, hence $k=\frac{d}{1+e^\epsilon}$. On the other hand, denote the derivative of $f'(k)$ as $f''(k)$, for any $\epsilon > 0.0, k \in (0,d)$, we have $f''(k)\geq 0$. Thus, $k=\frac{d}{1+e^\epsilon}$ minimizes $f(k)$. Further restricts $k$ to an integer value between $[0,d]$, Theorem \ref{the:l2k} holds.

\section{Proof of Theorem \ref{the:l2koptimal}}\label{app:l2koptimal}
Since mechanism $Q$ is symmetric for any $X_j\in \mathcal{X}$, we simply denote $g$ as $\sum_{Z\ni X_j} Q(Z|X_j)$, which is the hit rate of $X_j$ when secret value is $X_j$, denote $h$ as $\sum_{Z\ni X_j} Q(Z|X_j')$, which is the hit rate of $X_{j}$ when secret value is $X_j'$ ($j\neq j'$). We also denote $P_k=\sum_{|Z|=k} Q(Z|X_j)$, which is the probability of outputting a size $k$ subset in the mechanism.

Define $g'_k= \frac{k C_k}{k e^\epsilon+d-k}$, $h'_k=\frac{k C_k(k-1)+(d-k)k}{(k e^\epsilon+d-k)(d-1)}$, we have $g=\sum_{k=0}^d P_k g'_k$, $h=\sum_{k=0}^d P_k h'_k$. Note that $E[|\hat{\theta}-\theta|_2^2]=\frac{g(1-g)+h(1-h)d}{(g-h)^2}$, the following inequation holds.
\begin{equation}\label{eq:gmin}
E[|\hat{\theta}-\theta|_2^2]\geq \text{min}_{k=1}^d \frac{g'_k(1-g'_k)+h'_k(1-h'_k)d}{(g'_k-h'_k)^2}.
\end{equation}

Denote $f(g,h)=\frac{g(1-g)+h(1-h)d}{(g-h)^2}$, to prove (\ref{eq:gmin}), it is enough to show that $f(p g+ (1-p) g', p h+ (1-p) h')\geq \text{min}\{f(g,h), f(g',h')\}$ holds for any $0.0\leq g,h,p \leq 1.0$. Actually, since $2xy\leq x^2+y^2$,we have:
\begin{equation*}
\begin{aligned}
&f(p g+ (1-p) g', p h+ (1-p) h')&\\
&=\frac{(p g+ (1-p) g')-(p g+ (1-p) g')^2+d(p h+ (1-p) h')-d(p h+ (1-p) h')^2}{(p (g-h)+ (1-p)(g'-h'))^2}&\\
&\geq \frac{p (g(1-g)+h(1-h)d)+(1-p)(g'(1-g')+h'(1-h')d)}{(p (g-h)+ (1-p)(g'-h'))^2}&\\
&\geq \frac{p (g(1-g)+h(1-h)d)\,\,+\,\,(1-p)(g'(1-g')+h'(1-h')d)}{p (g-h)^2\,\,\,\,\,\,\,\,+\,\,\,\,\,\,\,\,\,\,\,\,(1-p)(g'-h')^2 }&\\
&\geq \text{min}\{f(g,h), f(g',h')\}.&\\
\end{aligned}
\end{equation*}

As a result of (\ref{eq:gmin}), Theorem \ref{the:l2k} and  $1.0 \leq C_{|Z|}\leq e^\epsilon$, which produces $f(g'_k,h'_k)\geq f(g_k, h_k)$ where $g_k=\frac{k e^\epsilon}{k e^\epsilon +d-k}$ and $h_k = \frac{k e^\epsilon}{k e^\epsilon +d-k}\cdot \frac{k-1}{d-1}+\frac{d-k}{k e^\epsilon +d-k}\cdot \frac{k}{d-1}$, Theorem \ref{the:l2koptimal} holds.

\end{document}